\documentclass[twocolumn,showpacs,preprintnumbers,amsmath,amssymb]{revtex4}
\usepackage{graphicx}
\usepackage{dcolumn}
\usepackage{bm}

\input epsf

\begin{document}

\title{A solvable senescence model showing a mortality plateau}
\author{J.~B.~Coe$^1$, Y.~Mao$^1$, M.~E.~Cates$^2$}
\affiliation{
$^1$Cavendish Laboratory, Madingley Road, Cambridge, CB3 0HE, UK \\
$^2$Dept.~of Physics and Astronomy, University of Edinburgh,\\ 
King's Buildings, Mayfield Road, Edinburgh, EH9 3JZ, UK
}
\date{\today}


\begin{abstract}
We present some analytic results for the steady states of the Penna model of senescence, generalised to allow genetically identical individuals to die at different ages via an arbitrary survival function. Modelling this with a Fermi function (of modest width) we obtain a clear mortality plateau late in life: something that has so far eluded explanation within such mutation
accumulation models. This suggests that factors causing variable mortality within genetically 
identical subpopulations, which include environmental effects, may be essential to understanding the mortality plateau seen in many real species. 
\end{abstract}

\maketitle





A common feature of the life-history of multicellular organisms is 
the progressive decline in the various physiological processes once the 
reproductive phase is complete. This process of aging has attracted 
considerable attention \cite{ageing,age2,rev,cw,Hwang}, particularly 
after work from Human Genome project identified specific `clock genes' \cite{clock}
which regulate human aging. In addition to these underlying genes 
the environment can play a part. For example, 
experimental studies \cite{longevity,long2} have shown that organisms 
subject to a reduction in caloric intake without essential nutrient deficiency 
display an extension of maximum life-expectancy. How, then, do the effects of 
evolution and/or the environment `pick out' a particular profile of senescence?

Two major theories of aging prevail, the `mutation accumulation' where mutations 
affecting old ages accumulate thanks to a weaker reproductive selection, and the 
`antagonistic pleiotropy' where a gene, beneficial in youth, is deleterious in old 
age \cite{cw}. 
Traditional mutation accumulation theories lead to an 
exponentially increasing mortality with age, the so-called Gompertz Law \cite{ageing}.
However, it has been experimentally observed that the rate of mortality, 
whilst obeying the Gompertz law up to an intermediate age, often shows
an unexpected drop at an advanced age. This drop, valid for many 
species from human to medflies, gives rise to a `mortality plateau', 
somewhat at odds with theory \cite{pla,pla2,pla3,pla4}.
As a result, `antagonistic pleiotropy' has been suggested as an essential part
of aging \cite{cw,sousa}. 

In 1995, with computer simulation in mind, 
Penna proposed a mutation accumulation model of senescence, 
which quickly became widespread \cite{penna}. Gompertz behaviour has
been reported for the standard Penna model \cite{penna2}, and variations on
the model were proposed to account for demographic features \cite{three,fermi,fermi2}, 
as well as catastrophic mortality \cite{salmon,exact}. 
However, a major shortcoming 
of the Penna model \cite{pennaflaw} was examined only recently: according to the model, 
all members of a genetically identical population must die exactly at the same age, 
which is obviously untrue.
Recent attempts in removing this determinism within the Penna model gave rise to a 
gently decelerating old age mortality, without reproducing a satisfactory 
`mortality plateau' \cite{huang}. 

Here we present a general formulation and an exact solution to the Penna model, 
which together with a simple Fermi survival function, gives rise to a convincing 
mortality plateau.
The introduction of a gene-dependent survival curve (replacing a single parameter, the 
deterministic age of death) enlarges the parameter space for simulations considerably. 
This makes analytical results all the more valuable, especially those that do not 
depend on the survival function chosen.
 
Below, we first consider a particular version of the Penna model
(by setting the mortality threshold $T=1$), and present for it an exact analytic solution.
Furthermore, this solution is robust against changes in the survival function. 
In the case of a Fermi function for survival, we show the presence of an 
extended `mortality plateau'. The generalisation of the analytic solution to $T>1$ is then given.

In the original Penna model \cite{penna}, 
each individual carries a string of binary numbers, which stays fixed for that
individual's lifetime.
A $1$ for the $i$-th bit represents the effect of a harmful gene which causes the individual to be struck by a heritable disease upon reaching the age of $i$.
Thus, as an individual ages, its bit-string is sequentially 
examined and diseases are accumulated. The individual expires 
upon encountering the $T$-th deleterious bit, where $T$ is a preset mortality 
threshold. Qualitatively, $T$ represents the number of heritable diseases required to make an individual nonviable; within the original model, if the $T$-th $1$ occurs at 
the $l$-th bit, the individual will survive precisely to age $l$ and no more.
As long as it survives, the
individual has a probability (rate) $b$ to give birth. The offspring inherits the same bit-string from the parent except for some mutations occuring at a 
small rate $\beta$. In simulations, the birth rate $b$ is often regulated by a 
`Verhulst factor' to prevent either population explosion or decay \cite{ageing}. 
A different Verhulst factor, reducing populations of all ages at every time step
(but with a fixed birth rate), 
has also been used, for which the Penna model can be formulated analytically into 
an eigenvalue equation and solved numerically \cite{alme}.

In the $T=1$ limit, the population can be described by a probability
function $n(l)$ of the first deleterious bit occuring at position $l$.
Equivalently $n(l)$ is simply the number of such $l$-type individuals at any 
particular time. 
Then the evolution of $n(l)$ is given by
\begin{equation}
\frac{d n(l)}{dt} = b n(l) e^{- \beta l} - \frac{n(l)}{l} + 
b (1-e^{-\beta})  e^{- \beta l} \sum_{l'=l+1}^\infty n(l')
\label{eq:one}
\end{equation}
where the first term corresponds to the mutation-free reproduction; the second
term gives the mortality, where $1/l$ is the average rate of mortality for an $l$-type
(which assumes the population changes slowly in the lifetime of an individual);
the third term represent the switching from $l'$ to $l$ due to a single 
mutation \cite{note}. Note that there is no noise in this equation. Hence it describes the thermodynamic limit of a large population in which deterministic dynamics is recovered.
Note also that in a stationary state of the original Penna model, $l$-types are distributed uniformly over all ages up to $l$ at which they promptly die, so the mortality rate is exactly $1/l$. However, our formulation is more general in that $l$-types may have an arbitrary survival function provided their mortality rate averages to $1/l$. (This is the same as demanding the average life expectancy for an $l$-type to be $l$, which we may take as the definition of $l$.) As in the original Penna model, only bad mutations (from $0$ to $1$) have been included. 

For a stationary state, the constant introduction of bad mutations is balanced by the longer reproductive lifetime of healthier individuals.
Setting $dn/dt=0$,
Eq.(\ref{eq:one}) can be solved exactly to give the recursion relation:
\begin{equation}
\frac{n(l+1)}{n(l)}=\frac{l+1}{l} \; \frac{ e^{\beta l} - b l}{ 
e^{\beta (l+1)} - b (l+1) e^{-\beta}}
\label{recur}
\end{equation}
Before going further, we need to examine the inter-dependence of 
the mutation and birth rates $\beta$ and $b$. 
Obviously a very large $b$ coupled with a negligible $\beta$
would result in an exponential population growth. A stationary 
state requires a sepcific combination of $\beta$ and $b$.
An extremely long-lived individual (large $l$) producing many offspring
in total (proportional to $l$), but mutation reduces the 
probability of it reproducing itself accurately by an exponential factor ($e^{-\beta l}$). 
This means that a very large $l$ cannot maintain itself, and must rely 
instead on mutated reproduction of even fitter individuals. 
Such a cascade is not possible in a finite population which must, 
at any time, contain a maximum $l = l_{max}$; the above argument shows that, if too 
large initially, this will decrease with time until the $l_{max}$ subpopulation is 
indeed self sustaining. 
The evolution equation Eq.(\ref{eq:one}) for this subpopulation then reads:
\begin{equation}
\frac{d n(l_{max})}{dt} = g(l_{max}) \; n(l_{max})
\end{equation}
where the growth rate function $g(l) \equiv b e^{-\beta l} - 1/l$,
and a steady state demands $g(l_{max}) = 0$, namely:
\begin{equation}
b e^{-\beta l_{max}} - 1/l_{max} =0.
\label{fixlmax}
\end{equation}
Next we argue that for a steady state in the thermodynamic limit, we should have:
\begin{equation}
g(l_{max}-1) < 0; \quad \quad g(l_{max}+1) \le 0.
\label{derivative}
\end{equation}
Without the first condition,
the subpopulation with $l= l_{max}-1$ would grow thanks to the extra proliferation
of $l_{max}-1$ as a result of a mutation to $l_{max}$. The second condition
ensures the stability of $l_{max}+1$ upon approaching steady state.
In the continuum limit, this is equivalent to demanding that
the derivative of $g(l)$ should be zero.


These conditions are well confirmed by our numerical simulations of the Penna model. However, if $n(l_{max}+1)$ falls to zero discontinously significantly before the steady state is reached, then the second condition in Eq.(\ref{derivative}) might not be strictly required. This is unlikely in the thermodynamic limit and we ignore such contingencies from now on. It would also happen if a maximum bit-string length less than $l_{max}$ were imposed in a simulation; see below.

Equations (\ref{fixlmax},\ref{derivative}) together lead to:
\begin{equation}
\frac{1}{e^\beta-1} \le l_{max} < \frac{1}{e^\beta-1}+1
\end{equation}
which implies a unique integer value for $l_{max}$ given any $\beta$.
Whenever $1/\beta$ is an integer, it lies in the above specified range, and 
Eqs.(\ref{fixlmax},\ref{derivative}) greatly simplify to:
\begin{equation}
l_{max}=1/\beta; \quad \quad b=\beta e
\label{stationc}
\end{equation}
where only one of $l_{max}$, $\beta$ and $b$ is a free variable. Choosing $\beta$ and $b$ such that $l_{max}=20, 30$, the result of the 
recursion relation Eq.(\ref{recur}) is plotted in Fig.1, with normalisation $\sum_l n(l)=1$. 

\begin{figure}[t!]
  \begin{center}
    \leavevmode
    \epsfxsize=7.5cm
    \epsfbox{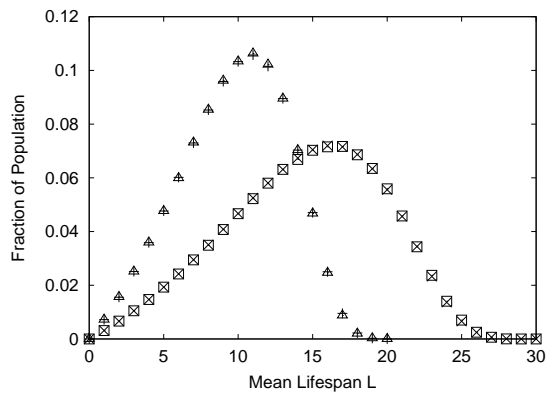}
\caption{Lifespan distribution $n(l)$ for $l_{max}=20 \;(+),\; 30 \;(\times)$,
compared with simulations (boxes). Simulation size $10^7$, averaged over $10$ runs.}
\vspace*{-5mm}
  \label{fig1}
  \end{center}
\end{figure}

Our simulation results are obtained following the method proposed 
by Penna \cite{penna}, where mutation rate is  pre-specified
as is the maximum string length $l_{sim}$ (often set to $32$). However,
a Verhulst factor is included to regulate the birth reproduction rate and 
the system finds its own stationarity by adjusting the birth rate. 
The string length dependence has been investigated by computer simulation, but no universal 
results were found \cite{stringlength}. Our analysis suggests that the results
depend on the value of $l_{sim}$ in relation to $l_{max}$ which is set by
the mutation rate, Eq.(\ref{stationc}). If $l_{sim} > l_{max}$, no
string length effect is expected as the final bits of the string do not
affect the system. Therefore, the most efficient simulation
is performed when $l_{max}$ approaches $l_{sim}$ from below.

With the original Penna model,
the only mortality at age $x$ occurs to individuals with $l=x$; 
these have number density $n(l)/l$ and mortality rate $1$. 
In other words, the survival 
function of an $l$-type is a square function with a height of $1$ and 
width $l$.
The normalised mortality at $x$ is therefore:
\begin{equation}
{\cal M} (x) = \frac{n(x)/x}{\sum_{l=x}^\infty   n(l)/l}
\end{equation}
where the denominator gives the total number of individuals living
at the age $x$, and the numerator gives the number of individuals dying
at the age $x$. The resulting population mortality exhibits a Gompertz-like behaviour up to an
intermediate age but then slows down as $l_{max}$ is approached \cite{penna2}, 
see Fig.2. But since the mortality at this point reaches one, there is no population left to continue the incipient plateau and only its onset is observed.

\begin{figure}[t!]
  \begin{center}
    \leavevmode
    \epsfxsize=6.5cm
    \epsfbox{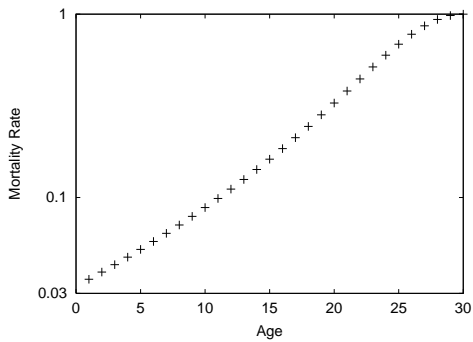}
\caption{Mortality rate with $l_{max}=30$, (note log scale).}
\vspace*{-5mm}
  \label{fig2}
  \end{center}
\end{figure}

However, equations (\ref{recur}-\ref{stationc}) make no assumption about the
survival function of an $l$-type and apply equally well to, say, 
a Fermi function \cite{fermi,fermi2}:
$$
f(x,l)=\frac{{\cal N}}{1 + \exp \frac{x-\tilde l}{w \tilde l}}
$$
where $w$ is its width in units of the `Fermi level' $\tilde l$. ${\cal N}$ ensures 
$f(0,\tilde l)=1$, and $\tilde l$ is chosen so that the average life span 
$\sum_x f(x,l) = l$. In fact, $\tilde l = l$ to within terms of order $\exp[-1/w]$; the latter are safely ignored in following plots, where we choose small
$w$'s as examples of modest influence of environmental factors on the life 
expectancy of genetically identical individuals. 
The analysis leading to the recursion result Eq.(\ref{recur})
stands, and the mortality function corresponding to Fig.2 is plotted in Fig.3.

\begin{figure}[t!]
  \begin{center}
    \leavevmode
    \epsfxsize=6.5cm
    \epsfbox{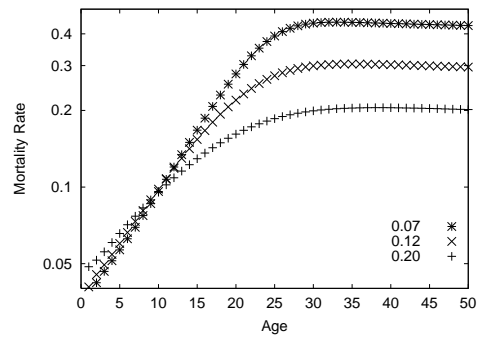}
\caption{Mortality rate for Fermi survival functions of different width $w$, 
and $l_{max}=30$.}
\vspace*{-5mm}
  \label{fig3}
  \end{center}
\end{figure}

A pronounced `mortality plateau' is now observed.
By varying $w$ in a reasonable range of $0.01-0.5$ and the choice of $b$,
which in turn determines $n(l)$, it is possible to obtain different
shapes of this `mortality plateau' which resemble those observed 
experimentally in various species \cite{pla}.
The different values of $w$ could be
easily justified as the species-dependent susceptibility to environmental 
variations as well as other factors. Thus, the Penna model, coupled
to a Fermi survival function, is sufficient to account for
the plateau observed in mortality for different species. 

\begin{figure}[t!]
  \begin{center}
    \leavevmode
    \epsfxsize=6.5cm
    \epsfbox{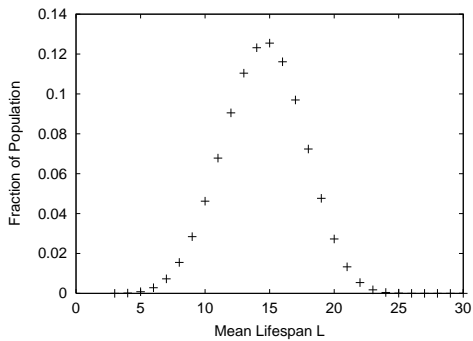}
\caption{Lifespan distribution function $C^l_{T-1} n_T(l)$ for $T=4$.}
\vspace*{-5mm}
  \label{figmany}
  \end{center}
\end{figure}

\begin{figure}[t!]
  \begin{center}
    \leavevmode
    \epsfxsize=6.5cm
    \epsfbox{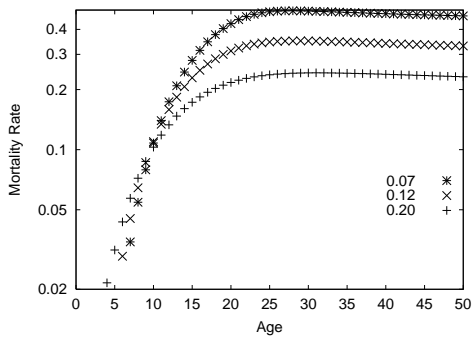}
\caption{Mortality rate for $T=4$.}
\vspace*{-5mm}
  \label{figmanymort}
  \end{center}
\end{figure}

Now we extend the analysis to the multi-disease cases where $T>1$. 
In this scenario, 
the obvious extension of our existing theory is to write 
$n(l_1,l_2,...,l_T)$ which gives the number of individuals with
deleterious bits at positions $l_1,l_2,...,l_T$ on their `genetic' 
strings. And a similar, albeit more complex, evolution equation can
be constructed. However, it is immediately 
evident that the final $l_T$ holds a special position as 
this is the bit which determines the individual's death. 
In contrast, the positions of other deleterious bits are 
less significant, in fact so insignificant as to inspire an
ansatz: 
$$
n_T(l)=n(l_1,l_2,...,l)
$$
for the steady state. 
Neglecting multiple mutations which will be addressed elsewhere \cite{future}, 
we find that the relevant generalization of Eq.(\ref{eq:one}) reduces to:
\begin{eqnarray}
0 &=&\frac{d n_T(l)}{dt} = b n_T(l) e^{- \beta (l-T+1)} - \frac{n_T(l)}{l} 
\label{ss}
\\
& &+ bT (1-e^{-\beta})  e^{- \beta (l-T+1)} \sum_{l'=l+1}^\infty n_T(l').
\nonumber
\end{eqnarray}
Our notation means $n_1(l) \equiv n(l)$, defined earlier.
The ansatz implies that the distribution function $n(l_1,l_2,...,l)$
depends merely on the position $l$ and not on any of the others. 
It is possible that there are other solutions not contained in the ansatz; 
without claiming uniqueness we note that our simulations show the ansatz leads to
a valid solution. 
Eq.(\ref{ss}) can be solved to give the recursion:
$$
\frac{n_T(l+1)}{n_T(l)}=\frac{l+1}{l} \; \frac{ e^{\beta (l-T+1)} - b l} 
{e^{\beta (l-T+2)} - b (l+1) (1-T+Te^{-\beta})}
$$
Since for every $l$ there are different combinations of having the
remaining $T-1$ mutations, the correct normalisation is now
$\sum_l  C^l_{T-1} \; n_T(l) =1$,
where $C^l_{T-1}$ is the number of combinations of choosing $T-1$ out
of a total of $l$, and $C^l_{T-1} \; n_T(l)$ gives the weighted
probability of finding an $l$-type.
The stability criterion for the multi-disease case reads:
\begin{equation}
l_{max}=1/\beta; \quad \quad b = \beta \;e^{1-\beta(T-1)} 
\end{equation}
which recovers Eq.(\ref{stationc}) for $T=1$.
Choosing $l_{max}=30$ as before and $T=4$, 
we plot the weighted mutation distribution function 
in Fig.\ref{figmany}. 
The mortality rate with the same Fermi survival function as before is
presented in Fig.\ref{figmanymort}. 

In conclusion, we have analysed the $T=1$
case of the Penna model, finding a nontrivial steady state solution for an arbitrary $l$-dependent survival function, and gained by ansatz similar solutions for the general $T$ case. 
Our result suggests the mortality plateau arises from a variation in the mortality within genetically identical subpopulation.
Our analysis also shows the importance of the maximum sustainable longevity $l_{max}$, 
set by the mutation rate, which determines for example whether there is any effect of 
the bit-string length $l_{sim}$ used in simulations (there is such an effect only if 
$l_{sim} < l_{max}$). It is hoped that the analytic solution presented here will provide 
guidance for further simulations. 
In our own future work we will describe the continuum limit (in $l$) of the theory 
presented here 
and deal with other, more technical aspects of the Penna model
such as multiple mutations and the Verhulst factor coupled with noise \cite{future}.

\end{document}